\documentclass[aps,prd,twocolumn,eqsecnum,showpacs,amsmath]{revtex4}
\usepackage{amsmath,amsfonts,amssymb}
\usepackage{dcolumn}
\usepackage{graphics}
\usepackage[dvips]{graphicx}
\newcommand{\abs}[1]{\left\lvert #1 \right\rvert}
\newcommand{\pd}{\partial}
\pacs{04.25.dg; 04.50.+h; 04.50.Gh; 04.50.Kd; 04.70.-s; 04.70.Dy; 04.80.-y}
\begin{document}

\title{Maeda-Dadhich Solutions as Real Black Holes}

\author{S.O. Alexeyev}
 \email{alexeyev@sai.msu.ru}
 \affiliation{%
Sternberg Astronomical Institute, Lomonosov Moscow State University, Universitetsky Prospekt, 13, Moscow 119991, Russia}%

\author{A.N. Petrov}
 \email{alex.petrov55@gmail.com}
 \affiliation{%
Sternberg Astronomical Institute, Lomonosov Moscow State University, Universitetsky Prospekt, 13, Moscow 119991, Russia}%

\author{B.N. Latosh}
 \email{latosh.boris@gmail.com}
 \affiliation{%
Faculty of Natural and Engineering Science, Dubna International University, Universitetskaya Str., 19, Dubna, Moscow Region, Russia;\\ Bogoliubov Laboratory of Theoretical Physics, Joint Institute for Nuclear Research, Joliot-Curie 6, 141980 Dubna, Moscow region, Russia}%

\date{\today}

\begin{abstract}
A four-dimensional static Schwarzschild-like solutions obtained in \cite{MaedaDadhich1, Maeda:2006hj, MaedaDadhich3, Molina:2008kh} in the frames of the Einstein-Gauss-Bonnet gravity at the Kaluza-Klein split are analyzed. In such models matter is created by auxiliary dimensions The main goal of our work is to check that these solutions are physically sensible, and to examine their characteristics, which could be observable. A non-contradictive definition of a total mass (energy) is given. Study of the perturbed equations demonstrates a possibility of their stability under linear perturbations. Depending on the combination of the parameters, black hole-like objects with one or two horizons or naked singularity are described in detail. Stable orbits of test particles around these black holes are presented. We show the exotic thermodynamical properties of the solution, in which the Hawking evaporation law has the behavior opposite to the one, usual in General Relativity. Unfortunately, current astronomical data does not allow to distinguish especial observable evidences, which we find for the solutions under consideration, from usual Schwarzschild ones.

\end{abstract}

\maketitle

\section{Introduction}
\label{Introduction}

String/M-theory \cite{SSM} is still a perspective candidate for the unified theory of all physical interactions, but a problem of its experimental verification remains unsolved. However, the generic formulation of a string theory is not completed and in some cases it is appropriate to consider the low energy approximation. The second order expansion is the Gauss-Bonnet one. In the framework of Einstein-Gauss-Bonnet (EGB) gravity a set of interesting solutions has been obtained, especially black hole ones \cite{EGB-BHs}. One can use the results of their physical effects to test the string theory and to search for the new physics in gravitation. In the present study we are looking for such effects in the solution obtained in the Maeda and Dadhich (MD) papers \cite{MaedaDadhich1}-\cite{MaedaDadhich3} devoted to EGB gravity with a cosmological constant. Of course, basic properties of the MD solutions were analyzed carefully by Maeda and Dadhich \cite{MaedaDadhich3}. On the other hand, properties which could be useful for experimental/observational physics were not considered completely. Thus, it is possible to study physically sensible properties and characteristics, which could be observable.

The main assumption for the MD solutions is the Kaluza-Klein type splitting of the EGB space-time. So, there is a physical (dynamic) space-time (of 2, 3 or 4 dimensions) and a space of auxiliary dimensions. The later one has a constant curvature with the negative sign, whereas in the standard Kaluza-Klein model a space-time of auxiliary dimensions is flat. We concentrate on the 4-dimensional (4D) static Schwarzschild-like solution because it seems to be more promising in searching for physically sensible effects. For simplicity we restrict ourselves to the minimal case of 6D EGB space-time that has been studied separately by Molina and Dadhich \cite{Molina:2008kh} and we denote as DM. The EGB solution properties in 6D are similar to those in arbitrary EGB dimensions with $N>6$.

We check the DM solutions from two points of view. To be physically sensible these solutions have to have, first, acceptable theoretical description, and, second, observable evidences, at least, principally. Theoretical description has to give a non-contradictive definitions of important characteristics. Undoubtedly, for solutions, which pretend to present astrophysical gravitating objects, such a characteristic is a total mass (energy) of the system. A separate section is devoted to this problem, where we use a well developed superpotential technique. Not less important theoretical property is a stability of the solution. We notice that a stability is directly connected with a well defined total mass, and independently we state a possibility that the DM solutions to leave stable. Thus, chances to detect related astrophysical objects arise.

Properties, which could be potentially observable and which we study are orbital and thermodynamic effects related to the DM solutions. The last part of the paper is just devoted to these problems. As known, thermodynamic properties are directly connected with a horizon structure of solutions. As it turns out, geometrical structure of the DM solutions is quite non-trivial and we examine it in a significantly more detail than it is given in \cite{MaedaDadhich1}-\cite{MaedaDadhich3}.

Our paper is organized as follows. In section \ref{DMmodel}, we briefly recall the MD idea of constructing solutions of specific type in the EGB gravity, present the DM solution and its main features. In section \ref{Thetotalmass}, we discuss the problems of the total energy in the DM models and present the total mass calculation. In section \ref{Stability} we study the stability of the DM solutions. In section \ref{Metric} we discuss geometrical structure of the DM solution in the form of black hole-like objects with one/two horizons or naked singularities. In section \ref{Orbits}, we establish a possibility of stable orbits and corresponding effects. In section \ref{Thermodynamics}, we calculate thermodynamical properties. Section \ref{DiscussionConclusions} contains a discussion of DM solution role and corresponding conclusions. In appendix we present differential operators for section \ref{Stability}.

\section{Dadhich-Molina solution}\label{DMmodel}

The restriction of the EGB gravity in 6D in the DM solution \cite{Molina:2008kh} leads to the following form of action:
\begin{equation}\label{I_action}
S=\frac{1}{2\kappa_6}\int d^6x \sqrt{-g}\left(R-2\Lambda+\alpha L_{GB}\right)+ S_{\rm matter}.
\end{equation}
Here and below we use $c=\hbar=1$ units, $\kappa_6$ is the 6-dimensional gravitational (Einstein) constant, $\alpha>0$ is the Gauss-Bonnet coupling constant; $R$ is a 6-dimensional Ricci scalar, $\Lambda$ is the cosmological constant and $L_{GB}$ is the Gauss-Bonnet term:
\begin{equation}
L_{GB}= R^2-4 R_{\mu\nu}R^{\mu\nu}+R_{\mu\nu\rho\sigma}R^{\mu\nu\rho\sigma} ~.
\end{equation}
Varying the action \eqref{I_action} with respect to metric, one obtains EGB gravitational equations:
\begin{equation}\label{I_gravity_eq}
{G}^\mu_{~~\nu} +\alpha {H}^\mu_{~~\nu} +\Lambda \delta^\mu_{~~\nu}=\kappa_6 {T}^\mu_{~~\nu}
\end{equation}
where $G_{\mu\nu}$ is the Einstein tensor:
\begin{equation}
G_{\mu\nu}=R_{\mu\nu}-\cfrac12\, g_{\mu\nu} R ,
\end{equation}
and $H_{\mu\nu}$ is the tensor corresponding to the Gauss-Bonnet term $L_{GB}$ in the action.

The main MD assumptions are as follows:
\begin{itemize}
\item The space-time of EGB gravity (6D in our case) is homeomorphical to $M^4\times K^2$ where $M^4$ is a 4-dimensional physical space-time and $K^2$ is a space of constant curvature with radius $r_0$ and sign $\bar k$ (they are not determined now). It is just the Kaluza-Klein type splitting.
\item EGB vacuum case $S_{\rm matter}=0$ is considered.
\end{itemize}

Under such assumptions the gravitational equations \eqref{I_gravity_eq} are automatically decomposed into the system:
\begin{eqnarray}
\left[1+\cfrac{4{\bar k}\alpha}{r_0^2}\right]\overset{(4)}{G}{}^A_{~B}+ \alpha \overset{(4)}{H}{}^A_{~B}+\left[\Lambda-\cfrac{\bar k}{r_0^2}\right] \overset{(4)}{\delta}{}^A_{~B}=0 ~, \label{I_4d_grav_eq} \\
\delta^a_{~~b}\left[-\frac12\overset{(4)}{R}+\Lambda -\cfrac{\alpha}{2} \overset{(4)}L_{GB}\right]=0 ~,\label{MainEQ_pre}
\end{eqnarray}
where $A,B =0,\ldots 3$ and $a,b=4, 5$ and index ``$(4)$'' denotes four-dimensional quantities. Because the tensor $\overset{(4)}{H}{}^A_{~B}$ vanishes in 4D space-time equations \eqref{I_4d_grav_eq} acquire the form of the vacuum Einstein equations with a modified cosmological constant, while equation \eqref{MainEQ_pre} plays a role of a constraint for the main equations only.

However, the authors of \cite{MaedaDadhich1}~-~\cite{MaedaDadhich3} claim that the system of \eqref{I_4d_grav_eq}-\eqref{MainEQ_pre} has no black hole-like solutions. Furthermore they suggested that the coefficients in square brackets in \eqref{I_4d_grav_eq} be put equal to zero . It turns out that in the case when the EGB-DM parameters are restricted by the equalities
\begin{equation}\label{fine_tuning}
\bar k /r_0^2=-1/4 \alpha = \Lambda  ~,
\end{equation}
instead of \eqref{I_4d_grav_eq} and \eqref{MainEQ_pre} one can consider only the scalar constraint \eqref{MainEQ_pre} that can be rewritten in a simplier form:
\begin{equation}\label{MainEQ}
\overset{(4)}{R}+\alpha \overset{(4)}{L}_{\text{GB}} +\cfrac{1}{2\alpha} =0.
\end{equation}
The equation \eqref{MainEQ} has a set of interesting solutions. Among them there is the DM one, that we consider in this paper. It can be represented as
\begin{eqnarray}\label{DM_metric}
ds^2 = f_{\pm} dt^2 - \cfrac{dr^2}{f_{\pm}}-r^2 ( d \theta^2 - \sin^2\theta d \varphi) ~, \\
f_{\pm}= 1 +\cfrac{r^2}{4\alpha} \left[ 1 \pm \sqrt{\cfrac23 + 16 \left( \cfrac{\alpha^{\frac32} M}{r^3} - \cfrac{\alpha^2 q}{r^4} \right)} \right] ~,\label{f_pm_SI_unit}
\end{eqnarray}
where the integration constants $M$ and $q$ appear to be the main solution parameters. Historically, $M$ is treated as a mass of a black hole and $q$ is an additional charge.

\section{Total Mass}
\label{Thetotalmass}

Energy connected characteristics play crucial role in a description of potentially observable physical objects. So, first of all, we have to define and calculate total mass (energy) of the solution \eqref{DM_metric}-\eqref{f_pm_SI_unit}. Other conserved characteristic are zero by symmetries of the solution.

One finds that the 4-dimensional metric \eqref{DM_metric}-\eqref{f_pm_SI_unit} at $r \rightarrow \infty$ behaves as:
\begin{equation}
f_{\pm} \sim \frac{r^2}{l^2_{\pm}} +1 \pm \frac{2M_{\rm eff}}{r} \mp  \frac{q_{\rm eff}}{r^2} + O\left( \frac{1}{r^4} \right)\,.
\label{1}
\end{equation}
Asymptotically the solution \eqref{DM_metric}-\eqref{f_pm_SI_unit} tends to AdS metric
\begin{equation}
d\bar s^2 = \bar f_{\pm}dt^2 - \frac{dr^2}{\bar f_{\pm}}-r^2d\Omega^2;\qquad \bar f_{\pm}= \frac{r^2}{l^2_{\pm}} +1.
\label{2}
\end{equation}
Here and below ``bar'' denotes a background quantity. So, in \eqref{1} and \eqref{2} the effective curvature radius of anti-de Sitter (AdS), ${l_{\pm}}$,  effective mass parameter, $M_{\rm eff}$, and effective charge parameter $q_{\rm eff}$,  are defined, respectively, as
\begin{eqnarray*}
{l^2_{\pm}}&=&\frac{12\alpha}{3\pm \sqrt{6}},\\
M_{\rm eff}&=& \sqrt{\frac{3}{2}}\alpha^{1/2}M\\
q_{\rm eff}&=& \sqrt{6}\alpha q.
\end{eqnarray*}

So, the DM solution \eqref{DM_metric}-\eqref{f_pm_SI_unit} behaving as (\ref{1}) can be considered as a perturbed metric respectively (\ref{2}). Such a presentation allows using the superpotential technique for a total mass calculation. The asymptotic metric is chosen as a background space-time. It could be written in two ways: both in 4 physical dimensions and in full 6 EGB gravity dimensions. Thus, the background metric could be chosen in the form (\ref{2}) as $\bar g_{AB}$, or (\ref{2}) together with the metric of $K^2$ space-time, where $\bar g_{\mu\nu}=\bar g_{AB}\times \bar g_{ab}$ ($\bar g_{ab}= g_{ab}$). Such a background provides the following time-like Killing vector:
\begin{equation}
 \xi^\alpha = (1, {\bf 0})
 \label{3}
 \end{equation}
necessary for defining the energy; ${\bf 0}$ includes all the space-time dimensions.

Various forms of superpotentials in $D>4$ dimensions, ${\cal J}^{\alpha\beta}_D(\xi)$, applicable in AdS and AdS-like backgrounds, have been constructed both in general relativity (GR) and in EGB theory, see, for example, \cite{AbbottDeser82}-\cite{Petrov_Lompay_2013_a} and references there in. For future work it is useful to rewrite them in an united form via anti-symmetric tensors ${J}^{\alpha\beta}_D$ as
 $$
{\cal J}^{\alpha\beta}_D(\xi) = \frac{\sqrt{-\bar g_D}}{\kappa_D}J^{\alpha\beta}_D(\xi).
 $$
Here and below $g_D = {\rm det} g_{\mu\nu}$ and $\bar g_D = {\rm det} \bar g_{\mu\nu}$ are determinants of $D$-dimensional dynamical and background metrics respectively, $\kappa_D$ is the gravitational constant in $D$-dimensional space-time. Using a time-like Killing vector, one calculates a total mass in an D-dimensional space-time:
\begin{equation}
{\cal E}(\xi) = \lim_{r \rightarrow \infty}\frac{1}{\kappa_D}\oint_{\partial\Sigma} ds_i \sqrt{-\bar g_D}{J}^{0i}_D(\xi)
 \label{4}
 \end{equation}
where $\partial\Sigma$ is $(D-2)$-dimensional boundary of a $(D-1)$-dimensional space-like hyper-surface $\Sigma$ defined as $x^0 = \rm const$, $ds_i$ is the integration element on $\partial\Sigma$.

The discussed metric \eqref{DM_metric} with the asymptote (\ref{1}) looks very similar to the Reisner-Nordstr\"om-AdS solution:
\begin{equation}
f = \frac{r^2}{l^2} +1 - \frac{2M}{r} +  \frac{Q^2}{r^2}.
\label{5}
\end{equation}
Then it seems total mass of the solution \eqref{DM_metric}-\eqref{f_pm_SI_unit} could be calculated by the standard methods like the mass in (\ref{5}). However, the solution (\ref{5}) is a solution of GR equations, whereas \eqref{DM_metric}-\eqref{f_pm_SI_unit} resolves the EGB ones. Moreover, they degenerate into the simplest scalar form \eqref{MainEQ}. Thus, for applying the usual GR methods it is important to present a more serious foundation and we provide it below.

It is better to apply the generic formula (\ref{4}) to the EGB model \eqref{I_action} - \eqref{I_gravity_eq} using Killing vector (\ref{3}) from the very beginning. So, we take a Belinfante corrected superpotential constructed in \cite{Petrov2009} as:
\begin{equation}
{\cal J}^{\alpha\beta}_{D} =  {\cal J}^{\alpha\beta}_{E} +{\cal J}^{\alpha\beta}_{GB}.
\label{6}
 \end{equation}
Other superpotentials in \cite{AbbottDeser82}-\cite{Petrov_Lompay_2013_a} lead to the same result but through a longer path. The GR and the Gauss-Bonnet parts of the superpotential \eqref{6} are
\begin{widetext}
\begin{eqnarray}
{\cal J}^{\alpha\beta}_{E}&=&
 \frac{\sqrt{-\bar g_D}}{\kappa_D}\left( \xi^{[\alpha} \bar\nabla_\lambda h^{\beta]\lambda}
-\bar\nabla^{[\alpha}h^{\beta]}_{\sigma}\xi^\sigma  +h^{\lambda[\alpha}\bar\nabla_\lambda \xi^{\beta]}\right),
\label{7}\\
 {\cal J}^{\alpha\beta}_{GB} &=&
 {\jmath^{\alpha\beta}_{GB}} - \bar{\jmath}^{\alpha\beta}_{GB};
\label{8}\\
\jmath^{\alpha\beta}_{GB}  &= & {\alpha\over
\kappa_D}\bar\nabla_\lambda\left\{\sqrt{-g} R_\sigma{}^{\lambda\alpha\beta} +
 4 \sqrt{-g}g^{\lambda[\alpha}R^{\beta]}_\sigma
 +2\sqrt{-g}\left[R_\tau{}^{\rho\lambda[\alpha}
-R^{\rho\lambda}{}_\tau{}^{[\alpha} - 2 R^\lambda_\tau
g^{\rho[\alpha}\right.\right.
 \nonumber\\&+& \left.\left.2R^\rho_\tau g^{\lambda[\alpha} +2 g^{\rho\lambda} R^{[\alpha}_\tau +
R\left( \delta^\lambda_\tau g^{\rho[\alpha}- \delta^\rho_\tau
g^{\lambda[\alpha}\right)\right]\bar g^{\beta]\tau}\bar g_{\rho\sigma} \right\}\xi^\sigma-{2\alpha\sqrt{-g}\over \kappa_D}\left\{{ R}_\sigma{}^{\lambda\alpha\beta} +
 4 {g^{\lambda[\alpha}R^{\beta]}_\sigma} + \delta_\sigma^{[\alpha}
 g^{\beta]\lambda}R \right\}
 \bar\nabla_\lambda \xi^\sigma\,
 \nonumber
 \end{eqnarray}
\end{widetext}
where the metric perturbation $h^{\alpha\beta}$ is defined by the relation
\begin{equation*}
\sqrt{-\bar g_D} h^{\alpha\beta} = \sqrt{-g_D} g^{\alpha\beta}
 - \sqrt{-\bar g_D}{\bar g}^{\alpha\beta},
\end{equation*}
$\bar\nabla_\lambda$ is a covariant derivative constructed with the help of ${\bar g}_{\alpha\beta}$. It is important to notice that the GR part, ${\cal J}^{\alpha\beta}_{E}$ in (\ref{7}), being constructed in arbitrary $D$ dimensions, has the form of Belinfante corrected superpotential for GR \cite{PK}. In Minkowski background with Cartesian coordinates and translation Killing vectors ${\cal J}^{\alpha\beta}_{E}$ transforms into the well-known Papapetrou superpotential \cite{Papapetrou48}.

Now we turn to Eq. (\ref{4}). Taking into account the MD restrictions in EGB gravity from the section \ref{DMmodel}, one immediately concludes that $J^{0i}_D$ (when $D=6$)  depends upon physical metric $g_{AB}$ and its derivatives only in the physical sector. Besides,  $\sqrt{-\bar g_D} = \sqrt{-\bar g_{4}}\sqrt{\bar g_{D-4}}$ where $\bar g_{4} = {\rm det} \bar g_{AB}$ and $\bar g_{D-4} = {\rm det} \bar g_{ab}$. Further, because of a spherical symmetry in the metric discussed one needs a $01$-component of the superpotential, $x^1 = r$, $J^{0i}_D \rightarrow J^{01}_D$ and $\sqrt{-\bar g_{4}} = r^2\sin\theta$ for (\ref{2}). Thus, (\ref{4}) could be rewritten in the form:
 \begin{equation}
{\cal E}(\xi) = \lim_{r \rightarrow \infty}\frac{1}{\kappa_D}\oint d\theta d\phi(r^2\sin\theta) {J}^{01}_D(\xi) \oint_{r_0} dx^{D-4}  \sqrt{\bar g_{D-4}}.
 \label{9}
 \end{equation}

Returning to Kaluza-Klein idea from MD papers, one treats the additional dimensions as compact ones. Thus, a volume
 \begin{equation}
 V_2 = \oint_{r_0} dx^{D-4}  \sqrt{\bar g_{D-4}}
 \label{10}
 \end{equation}
  is finite, and Eq. (\ref{9}) is established as
 \begin{equation}
{\cal E}(\xi) = \lim_{r \rightarrow \infty}\frac{1}{\kappa_4}\oint d\theta d\phi(r^2\sin\theta) {J}^{01}_6(\xi).
 \label{11}
 \end{equation}
Here, according to the Kaluza-Klein prescription, $\kappa_4 = \kappa_6/V_2$ is the gravitational constant of a 4-dimensional space-time. So, one can conclude that Eq. (\ref{11}) represents the result of a gravitational interaction in a 4-dimensional space-time.
However, then one points out that the Eq. (\ref{11}) is incomplete because a superpotential ${\cal J}^{\alpha\beta}_6(\xi)$ has to be exchanged by ${\cal J}^{\alpha\beta}_4(\xi)$ constructed in 4 dimensions. But, a 4D metric theory, in the framework of which ${\cal J}^{\alpha\beta}_4(\xi)$ could be constructed, is not determined at the moment, although from the theoretical point of view GR one is more preferable. Moreover, it turns out that GR is preferable for applications. We test the solution \eqref{DM_metric}-\eqref{f_pm_SI_unit} on astronomical and cosmological scales where GR explains the experimental data correctly. Thus, GR is better to choose ${\cal J}^{\alpha\beta}_4(\xi)$ constructed in the framework of GR. So, we use the GR part (\ref{7}) of (\ref{6}) changing $\kappa_D \rightarrow \kappa_4$. Finally, instead of (\ref{11}) one has to use
\begin{equation}
{\cal E}(\xi) = \lim_{r \rightarrow \infty}\frac{1}{\kappa_4}\oint d\theta d\phi(r^2\sin\theta) {J}^{01}_E(\xi).
 \label{12}
 \end{equation}

The GR equations corresponding to (\ref{12}) with the solution \eqref{DM_metric}-\eqref{f_pm_SI_unit} can be derived as
\begin{equation}
G_{AB} + g_{AB}\Lambda_{\pm} = \kappa_4 {\cal T}_{AB}.
\label{13}
\end{equation}
Here $\Lambda_{\pm} = -1/l^2_{\pm}$. Of course, a trace of Eq. (\eqref{13}) must be compatible with the main equation \eqref{MainEQ}. Such a model based of the equations (\ref{13}) cannot be described in the framework of the standard GR. The Einstein tensor $G_{AB}$ is determined as usual, whereas the energy-momentum tensor ${\cal T}_{AB}$ of induced matter is generally unknown. Only one solution of the Eq. \eqref{MainEQ} determines a particular form of ${\cal T}_{AB}$ after the substitution of \eqref{DM_metric}-\eqref{f_pm_SI_unit} into the left hand side of  (\ref{13}). Therefore, we treat the theory based on the equations (\ref{13}) as an ``effective'' version of GR. Nevertheless, such a derivation is enough to define a superpotential ${\cal J}^{\alpha\beta}_E(\xi)$ without ambiguities because for construction ${\cal J}^{\alpha\beta}_E(\xi)$ one needs a pure gravitational presentation without matter part. So, we use (\ref{12}) to calculate total mass of the system (object) presented by the solution \eqref{DM_metric}-\eqref{f_pm_SI_unit}. Taking into account the asymptotic behavior (\ref{1}) and the background metric (\ref{2}) one obtains
\begin{equation}
{\cal E} = \pm M_{\rm eff}.
\label{14}
\end{equation}
A system is physically sensible if its total energy is not negative ${\cal E}\ge 0$, therefore one must set $\pm M_{\rm eff} \geq 0$. Furthermore, the charge parameter $q$ does not contribute into ${\cal E}$, so we have an analogy with the Reisner-Nordsr\"om-AdS black hole (\ref{5}), where ${\cal E}=M$, and $Q^2$ does not contribute into ${\cal E}$.

Finally, a move from (\ref{9}) to (\ref{14}) supports the MD assertion \cite{MaedaDadhich1}-\cite{MaedaDadhich3}, so the matter is induced by additional dimensions. Starting from the vacuum system \eqref{I_4d_grav_eq} and \eqref{MainEQ_pre} we derive formula (\ref{9}), which is defined for the vacuum EGB model and step by step we obtain (\ref{14}) with the correctly defined total mass of the system. Unambiguously, the mass defined by the parameter $M$ is a characteristic of matter created by compactified additional dimensions. The charge parameter $q$ also describes the induced matter. This can be proved by changing the limit in (\ref{12}) from $r \rightarrow \infty$ to $r \rightarrow R_0 = {\rm const}$. So a mass inside a sphere with the radius $R_0$ is determined both by the parameters $M$ and $q$. Although $q$ does not contribute into the total mass like $Q^2$ in electrodynamics.

Concluding the section, we return to (\ref{9}) again. Let us assume that we suppress a desire to transfer to a metric theory in 4-dimensional space-time like (\ref{14}). In other words, lets restrict ourselves to a consideration of 6D EGB theory only. Then, in (\ref{9}), one must use the complete superpotential (\ref{6}) - (\ref{8}). After direct and cumbersome calculations one finds ${\cal E}=0$ for the solution \eqref{DM_metric}-\eqref{f_pm_SI_unit}. This result coincides with the conclusions of \cite{Cai+}, where the solution \eqref{DM_metric}-\eqref{f_pm_SI_unit} represents a particular case of a Lovelock gravity.

\section{Stability}
\label{Stability}

The problem of DM solution stability is important from a theoretical point of view and is crucial in cosmological and astrophysics applications. Usually, for such test one perturbs metric coefficients and checks their further behavior. Technically this is carried out by perturbing initial field equations and solving them.
However, to obtain the DM solution special restrictions and conditions have been used. Therefore, a study of the perturbation problem cannot be provided in the standard steps and requires a more attention.

 A first fundamental question  that arises is  as follows. Have we to consider a stability of the DM solution in the 6-dimensional EGB spacetime or in the 4-dimensional physical spacetime? Let's consider the system \eqref{I_4d_grav_eq}, \eqref{MainEQ_pre} together with \eqref{fine_tuning}. Following the the paper by Cai with coauthors \cite{Cai+} one finds that vanishing coefficients in the vacuum Einstein equations \eqref{I_4d_grav_eq} correspond to an infinite gravitational constant. So, a model presented by all the equations \eqref{I_4d_grav_eq} and \eqref{MainEQ_pre}, of course, with \eqref{fine_tuning} in 6 dimensions is fundamentally stable. Then, because \eqref{fine_tuning} have to be left unchanged one has a possibility to consider perturbations of the unique equation \eqref{MainEQ} in 4-dimensional spacetime only. Beside of that, there is another argument for considering the conditions \eqref{fine_tuning} as unperturbed. Assume that one disturbs all the equations in 6 dimensions of EGB spacetime and recall that the DM solution exists {\em only}, when the conditions \eqref{fine_tuning} are fulfilled. Then, destructing conditions \eqref{fine_tuning} even by extremely small perturbations, or even by time independent perturbations, one finds immediately that the DM solution does not exist at all. Thus, perturbations of a general kind in 6 dimensions make the stability problem of the DM solution senseless itself.

 Another problem in studying the stability of the DM solution appears because one has a possibility to analyze the {\em unique} equation \eqref{MainEQ}, whereas in a general case one has to study 10 perturbed metric functions. One could try to use the 10 equations \eqref{13}. However, in this case, it is impossible to apply the standard methods because a general form of \eqref{13} is unknown. Nevertheless, the idea of the ``effective'' GR with the equations \eqref{13} could be useful. So, Abbott and Deser in \cite{AbbottDeser82} proved the stability of an asymptotically AdS system in the framework of GR. They based on the statement that total energy of the system is positive for asymptotically AdS space-time. In fact, if we are sure of the validity of the Kaluza-Klein mechanism of matter creation and the positive total mass (energy) (\ref{14}), the Abbott-Deser arguments are sufficient to state that the DM solution is also stable.

In spite of the above conclusion is quite concrete, it was obtained by a not direct way. Therefore it is interesting and important to consider the problem being restricted only to the unique equation. It is important that we cannot solve the perturbation problem completely and classically. Analyzing \eqref{MainEQ} we work within the linear metric perturbations. Then, each of the perturbations within a linearized equation \eqref{MainEQ} can be considered as separate. In such a presentation, we obtain several equations for each perturbation. So, we have no possibility to study the perturbation problem directly and fully. But it is possible to find a behavior of perturbations that does not destroy the DM solution (among others, which could destroy it). If such a behaviour exists, we assert that a system might be stable. For our purposes this is enough, and we follow such strategy.

It is constructive to consider two cases: when (i) axial symmetry is applied, (ii) there are no symmetries. The case (i) means the existence of perturbations with axial symmetry, corresponding to black holes with small angular momentum. This case is quite important because the MD model excludes rotating black holes as it is stated in \cite{MaedaDadhich1}-\cite{MaedaDadhich3}. Further, the instability of the DM metric with axial perturbations means that the solution must be destroyed by a rotation. Such a conclusion could be fatal for cosmological and astrophysics applications where most of objects rotate. 

We turn to an axial case in general and use the expression in a generic form with a non-stationary axially symmetric metric suggested by Chandrasekhar \cite{chandrasekhar1998mathematical}. It has the form:
\begin{eqnarray}\label{metric_p}
ds^2 = e^{2 N} dt^2 - e^{2 \Psi} \left( d\varphi -  Q dr -  P d\theta -\Omega dt \right)^2- \nonumber \\
-e^{2 M} dr^2 - e^{2 \Lambda} d\theta^2 ~
\end{eqnarray}
where $N$, $M$, $\Psi$, $\Lambda$, $\Omega$, $P$ and $Q$ depend upon $t$, $r$ and $\theta$ (usual spherical coordinates). For the DM solution that is treated as background metric these functions acquire the form:
\begin{eqnarray}\label{CH_background_begin}
	N = \cfrac12 \ln f_\pm(r)~, \\
	M =-\cfrac12 \ln f_\pm(r)~, \\
	\Psi=\ln(r \sin \theta)~,\\
	\Lambda=\ln(r) ~, \\
	\Omega =0 ~, \\
	P=0 ~,\\
	Q=0 ~.\label{CH_background_end}
\end{eqnarray}
Perturbing the functions \eqref{CH_background_begin}-\eqref{CH_background_end} and introducing a perturbative parameter $\varepsilon$ one gets:
\begin{eqnarray}\label{CH_p_begin}
	N = \cfrac12 \ln f_\pm(r) + \varepsilon \nu~, \\
	M =-\cfrac12 \ln f_\pm(r) + \varepsilon \mu, \\
	\Psi=\ln(r \sin \theta) + \varepsilon\psi ~,\\
	\Lambda=\ln(r) + \varepsilon\lambda ~, \\
	\Omega =\varepsilon \omega ~, \\
	P=\varepsilon p ~,\\
	Q=\varepsilon q ~,\label{CH_p_end}
\end{eqnarray}
where functions $\mu$, $\nu$, $\psi$, $\lambda$, $\omega$, $p$ and $q$ depending upon $t$, $r$ and $\theta$ are treated as perturbations. Recalling our assumption for a linear approximation we neglect the terms, quadratic in $\varepsilon$ in \eqref{metric_p}, which appear after substituting \eqref{CH_p_begin}-\eqref{CH_p_end}, and derive $g_{\mu\nu}$ matrix in the form:
\begin{widetext}
\begin{equation}\label{metric_final}
g_{\mu\nu}=
\begin{pmatrix}
(1+2 \varepsilon \nu) f_\pm &0&0&  \varepsilon \omega r^{2} \sin^{2}\theta \\
0 & -(1+ 2 \varepsilon \mu) ~ \cfrac{1}{f_\pm} &0 &  \varepsilon q r^{2} \sin^{2} \theta \\
0&0& -(1+2\varepsilon \lambda) r^{2} &  \varepsilon p r^{2} \sin^{2} \theta  \\
 & & & \\
\varepsilon \omega r^{2} \sin^{2}\theta & \varepsilon q r^{2} \sin^{2}\theta &  \varepsilon p r^{2} \sin^{2}\theta & -(1+ 2 \varepsilon \psi) r^{2} \sin^{2} \theta
\end{pmatrix} + o(\varepsilon^2) ~.
\end{equation}
\end{widetext}
Substituting \eqref{metric_final} into the main DM equation \eqref{MainEQ}, leaving the linear in $\varepsilon$ terms only and considering each perturbation separately one obtains the following system
\begin{equation}\label{DO_view}
D^{(\zeta)} \zeta =0~,
\end{equation}
where the notation $\zeta$ is used for each perturbation: $\zeta= \mu, ~\nu, ~\psi, \ldots, ~q$, $D^{(\zeta)}$ is a differential operator for $\zeta$ perturbation. Eq. \eqref{DO_view} implies that we can consider an influence of the DM solutions of different kinds separately. A concrete form for the differential operators in \eqref{DO_view} is presented in Appendix.

 Of course, the representation of the unique equation for perturbations in the form of the system \eqref{DO_view} significantly simplifies the analysis. As a result, we lose various variants to find many of both stable and unstable possibilities for behaviour, which could follow from the unique equation. However, if we find a behaviour of the set of the functions $\zeta$ that leaves the MD solution stable then our task to find a possibility of a stability of the MD solution is achieved.

As $D^{(p)}$, $D^{(q)}$ and $D^{(\omega)}$ are not differential operators, the equations \eqref{DO_view} for $p$, $q$ and $\omega$ lead to vanishing of $p$, $q$ and $\omega$ values. Eqs. \eqref{DO_view} for all the other perturbations allow to apply the procedure of variables separation and pick out time, radial and angular parts for each one. We concentrate on $t$ dependency, because it is crucial for the stability study.

For $\nu$ perturbation, $D^{(\nu)}$ does not contain time derivatives, so, among other possibilities, we can consider $\nu$ perturbation as time independent only. For $\mu$ perturbation, a time dependent part of $D^{(\mu)}$  has a simple form and is satisfied by a {\em time dependence} $\exp[- t / \tau]$, where $\tau$ is a constant of variables' separation. There are no restrictions on a complex number $\tau$ from a radial part equation. Therefore to find the perturbation $\mu$ that vanishes exponentially in time one treats $\tau$ as a positive real number. Operators $D^{(\psi)}$ and $D^{(\lambda)}$ contain a more complicated wave-like operator $D_{(r t)}$. However, similarly for perturbations $\psi$ and $\lambda$  with $D^{(\psi)}$ and $D^{(\lambda)}$ operators a behavior $\exp[- t / \tau]$  with an arbitrary complex number $\tau$ is permitted also. Therefore, in general, wave and wave-like solutions are possible. However, again one treats $\tau$ as a positive real number and finds that $\psi$ and $\lambda$ vanish exponentially in time. Thus, based on the result that perturbations can die away exponentially, we assert that the DM solution has a possibility to remain stable.

Finally, let's discuss the case (ii) without space-time symmetries. Firstly, one has to consider 10 perturbations, secondly, all of them depend upon the coordinates $t$, $r$, $\theta$ and $\varphi$. In general, 
the study looks very cumbersome. To simplify it we apply the idea of the ``effective'' GR. Because of the general covariance of the Einstein equations \eqref{13}, there is a freedom in coordinate transformations at the level of perturbations. Using this possibility we set 4 of 10 perturbations equal to zero from the very beginning. Using the Chandrasekhar frame we derive a perturbed metric matrix as
\begin{widetext}
\begin{equation}\label{metric_p_gen}
g_{\mu\nu} = \begin{pmatrix}
(1+2 \varepsilon \nu) f_\pm(r) & 0 & 0 &  0 \\
0 & -(1+ 2 \varepsilon \mu) ~ \cfrac{1}{f_\pm(r)} &0 &  \varepsilon q r^{2} \sin^{2} \theta \\
0 & 0 & -(1+2\varepsilon \lambda) r^{2} &  \varepsilon p r^{2} \sin^{2} \theta  \\
 & & & \\
0 & \varepsilon q r^{2} \sin^{2}\theta &  \varepsilon p r^{2} \sin^{2}\theta & -(1+ 2 \varepsilon \psi) r^{2} \sin^{2} \theta
\end{pmatrix} + o(\varepsilon^2) ~,
\end{equation}
\end{widetext}
instead of \eqref{metric_final}. The study of \eqref{metric_p_gen} in analogy to analyzing \eqref{metric_final} with \eqref{DO_view} leads to the same conclusion: perturbations can die away exponentially in time, and, thus, the DM solution has a possibility to conserve a stability.

\section{Geometrical Structure}
\label{Metric}

As the DM solution is stable, we analyze its geometrical structure up to $r \rightarrow 0$. For the sake of simplicity hereafter we work within $c=\hbar=\alpha=1$ units. Then $f_\pm$ defined in \eqref{f_pm_SI_unit} is rewritten as
\begin{equation}\label{f_pm}
f_\pm(r)=1+\cfrac{r^2}{4 } \left[1 \pm \sqrt{\cfrac23 + 16\left( \cfrac{M}{r^3}-\cfrac{q}{r^4} \right) } ~ \right]  ~.
\end{equation}
We call $f_+$ and $f_-$ positive and negative branches of the solution, respectively.

First of all, there are some general restrictions on parameters $M$ and $q$. The requirement for a total mass (energy) of the system \eqref{14} to be non-negative leads to the conclusion that for the $f_+$ branch the mass parameter $M$ is non-positive, whereas for the $f_-$ branch it is non-negative. The positive branch is not valid for black hole-like objects describing because $f_+$ do not vanish and, so, a horizon could not be defined. The negative branch can describe black holes because $f_-$ is allowed to acquire zero's values. Both $f_+$ and $f_-$ could correspond to objects with naked singularities. We are interesting in potentially observable consequences.
Therefore, as Eq. \eqref{14} is valid for both branches, it can be applied to describe both black holes and naked singularities. To study the charge parameter $q$, we compare the asymptotic behavior \eqref{1} with the RN-AdS solution \eqref{5} in electrodynamics. In the later case the coefficient before $r^{-2}$ is positive, whereas in the case that we discuss in the paper coefficient before $r^{-2}$ can also be negative. This means that the charge parameter $q$ can describe the additional gravitational potential with the asymptotes analogous to the electric charge in \eqref{5} and a tidal charge of the Dadhich-Rezania solution \cite{Dadhich:2000am}. Such an exotic charge describes a bulk influence on a brane. In the DM solution additional dimensions are a part of the construction, so, it is natural to expect a similar behavior for the charge $q$.

For both branches the expression under square root must be non-negative. So, the range of definition of $M-q$ is restricted by inequality:
\begin{equation}\label{f_pm exist_1}
r^4 + 24 M r - 24 q \geq 0.
\end{equation}
We consider it as a limitation for $q$ parameter in both $M>0$ and $M<0$ cases.

Before studying structures of black hole-like objects with horizons and naked singularities we have to state few necessary relations arising from \eqref{f_pm exist_1}. We start from the negative branch with $M > 0$. Let us rewrite \eqref{f_pm exist_1} in  the form
\begin{equation}\label{q_cons}
q\leq \cfrac{r}{24} \left(r^3+24 M \right) ~.
\end{equation}

It is natural to require the existence $f_-$ everywhere in $r \geq 0$ area both in black hole-like and naked singularity cases. The limitation \eqref{q_cons} leads to the following condition: $q\leq 0$. However, when the area inside the horizon $r_H$ is not defined, we can continue to treat an external area as a physical one. So, instead of \eqref{q_cons} we consider the relation:
\begin{equation}\label{q_positive_condition}
q\leq \cfrac{r_H}{24} \left( r_H^3 + 24 M \right) .
\end{equation}
As a result, there are two possible cases: $r \geq 0$ and \eqref{q_positive_condition}. For the naked singularity case only the restriction $q\leq 0$ is allowed.
We turn to the positive branch where $M < 0$. It describes naked singularities. To do this we return to \eqref{q_cons}, considering the case $r \geq 0$ and $M < 0$. This leads to a simple limitation:
\begin{equation}\label{q_f+_constrain}
q\leq -\cfrac18 (6 \abs{M} )^\frac43.
\end{equation}
Note that this is a restriction for $q$ in the region $M<0$.

We turn to the conditions, under which the discussed solution becomes a black hole. We return to the negative branch $f_-$. Horizons are defined by the following condition: $f_- = 0$ which appears when the inequality
\begin{equation}\label{Horison conditions}
r^4 + 24 r^2 -48 M r +48 q + 48 \leq 0  ~,
\end{equation}
following from \eqref{f_pm}: $f_- \leq 0$ holds.

To clarify, it is enough to consider a simpler case $q=0$. Then, \eqref{Horison conditions} can be rewritten in the form:
\begin{equation}\label{Horison condition q=0}
r^4 + 24 r^2 +48 \leq 48 M r .
\end{equation}
Such a configuration becomes a black hole, when the positive mass parameter $M$ achieves the critical value $M_c$. The equation \eqref{Horison condition q=0} is 4th order. Its analytical solution for the expression $M_c$ exists and can be obtained, using the discriminant of \eqref{Horison condition q=0} resolvent
\begin{equation}\label{M_c}
M_c = \cfrac23 \sqrt{2(1+\sqrt{2})}.
\end{equation}
In fact, it defines a radius for an unique horizon $r_c \simeq 1.3$.
Fig. 1 illustrates the situations presented by the inequality \eqref{Horison condition q=0} derived for the $q=0$ case: the upper curve corresponds to the case without horizons (naked singularity), the middle curve corresponds to a single horizon, and the lower curve corresponds to the case with two horizons.

\begin{figure}{h}
\begin{center}
\includegraphics[width=80mm]{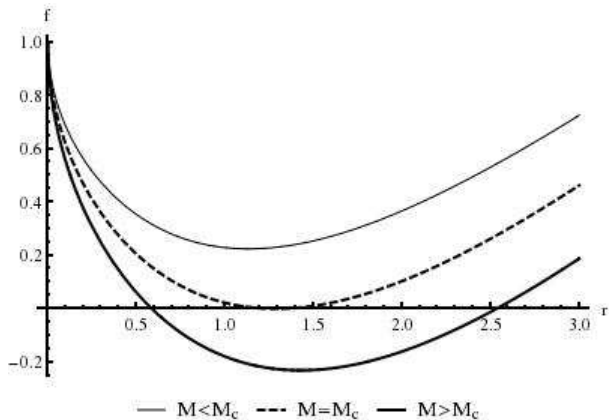}
\end{center}
\caption{$f_-$ functions for different masses}
\end{figure}

Here it is necessary to emphasise an interesting case with $M=0$ occurring for both branches $f_\pm$. As the $M$ parameter is coupled to the total mass \eqref{14} of the system, the $M=0$ case looks strange. To clarify, it is necessary to add the charge parameter $q$ to the consideration. Firstly, the meaning of $q$ differs from the electric charge $Q$ in electrodynamics. Indeed, $Q$ does not exist, when $M=0$, in other words, the Reisner-Nordstr\"om-AdS solution \eqref{5} is meaningless for  $M=0$ case. Secondly, as discussed after \eqref{14}, $q$ represents matter induced by additional dimensions. So, the condition $M=0$ restricts the consideration for the $f_-$ branch to $q\leq 0$ condition. 
Adding it to \eqref{q_f+_constrain} one finds that for $M=0$ the restriction $q\leq 0$ is fulfilled for both branches. Introducing the cutoff and, therefore, changing the integration limit from $r \rightarrow \infty$ to $r \rightarrow R_0 = {\rm const}$ in (\ref{12}), it appears to be possible to find a mass situated inside a sphere with the radius $R_0$ defined by $q$. Thus, the parameter $q$ totally defines the gravitational interaction in spite of the massless solution.
Such gravitational potential unlike the Newtonian's one has a fall-off behavior like $\sim r^{-2}$ analogously to the electric field one in (\ref{5}). However, there are no electrically charged objects in the Universe, because they would be immediately screened by opposite charged ones. From this point of view, a $q$-potential with the fall-off $\sim r^{-2}$ could be observed. For example, in combination with the usual Newtonian potential it could describe an irregular dynamics in galaxies. So instead of the idea of dark matter as new physics it is possible to treat it as geometrical contribution from string-inspired higher-dimensional space-time. This idea requires additional study.

Now we switch to a configuration with $M=0$ in the case of the $f_-$. Then, \eqref{f_pm} is rewritten as
\begin{equation}\label{f-M=0}
f_-=1+ \cfrac{r^2}{4}\left(1-\sqrt{\cfrac23 -\cfrac{16 q}{r^4}}\right)  ~.
\end{equation}
One obtains easily that for $q>-1$ there are no horizons, for $q\leq-1$ one horizon appears and there is no a possibility for two horizons. A possibility to construct horizons for totally massless objects looks quite surprisingly. Unlike GR, in the DM solution case the total mass determined by $M$ does not describe all the gravitational properties of the solution. This assertion supports our proposal about an analogy between the charge $q$ and a tidal charge from Dadhich-Rezania solution \cite{Dadhich:2000am} that describes bulk gravity influence at the brane. So, it turns out that gravitational effects related with $q$ are also induced by additional dimensions. Here we have to recall that a massless black holes described by one parameter in the Lovelock gravity are discussed in \cite{Cai+}.

Now, we present a generic picture with arbitrary values of $M$ and $q$. It is useful to rewrite expression \eqref{Horison conditions} in the following form:
\begin{equation}\label{Horison conditions analys}
r^4 + 24 r^2 + 48 \leq  48 M r - 48 q ~.
\end{equation}
Recall that it is a condition for existence of horizons, which exist in the case of the negative $f_-$ branch only. In the left hand side of \eqref{Horison conditions analys} there is a fourth order parabolic equation, while on the right hand side is linear one. If we draw plots of the parabola from left hand side of \eqref{Horison conditions analys} and the line from right hand side of \eqref{Horison conditions analys}, we see only three possibilities of mutual curves positions. First one is when line lies completely under parabola, and means an absence of horizons. Second case is when line crosses parabola at two points. Hence, there are two horizons. Lastly, when line is tangential to parabola, the equation in \eqref{Horison conditions analys} contains a single horizon only. These arguments show that there are three areas on $M$-$q$ parameters phase-space, corresponding to naked singularity configuration, one and two horizons configurations. Fig. 2 presents a corresponding diagram. To find the configurations with one horizon it is necessary to use the standard calculus. 
A ``one-horizon curve'' is defined by the following parameterization
\begin{equation}\label{one_horison_curve}
\begin{cases}
M= \cfrac{s^3}{12}+s ~,\\ \\
q =\cfrac{s^4}{16}+\cfrac{s^2}{2}-1 ~,
\end{cases}
\end{equation}
where $s$ is the point, at which the straight line touches the parabola. In the Fig. 2 it is drawn by a dashed curve on the right part of the picture. The condition \eqref{Horison conditions analys} shows that the area below the one-horizon curve is two-horizon configurations zone. Analogously the area above the one-horizon curve is to be the naked singularity configurations zone. However, it is so at $q > 0$. Indeed, if this naked singularity is allowed, we have to set $r \geq 0$, but then from \eqref{q_cons} the relation $q\leq 0$ follows only. Therefore, the area of naked singularity with a positive charge $q$ is forbidden, see Fig. 2. Only positive values of $q$ are restricted by \eqref{q_positive_condition}. To state the sense of the area under line $q=-1$ (more exactly $q\leq -1$) at the right part of the Fig. 2 one has to use the same arguments, which are used for the case $M=0$ and $q\leq -1$. One finds immediately that it is the one horizon zone.

To finish the discussion on Fig. 2, we return to the positive $F_+$ branch and use \eqref{q_f+_constrain}, that represents a curve at the left part of the picture, which divides it onto a forbidden zone and zone of naked singularity.

\begin{figure}[h]
\begin{center}
\includegraphics[width=80mm]{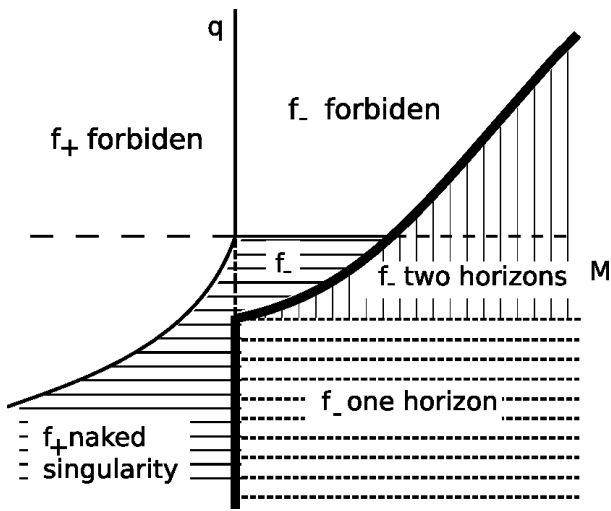}
\end{center}
\caption{Dadhich-Molina solution phase diagram}
\end{figure}
It seems to be interesting to compare the description of the DM objects with charged Schwarzschild or charged Schwarzschild-AdS solutions.
Everywhere the relations between two parameters, mass and charge, define whether there are two or one horizon or naked singularity configurations. The critical mass-charge relation defines the condition for one-horizon configuration existence that can be seen from Fig. 3. It is important to notice that configurations situated left from the Dadhich-Molina critical charge plot are forbidden. We also skip $q<0$ area, because for the charged solutions in GR this region is non-physical.

\begin{figure}[h]
\begin{center}
\includegraphics[width=80mm]{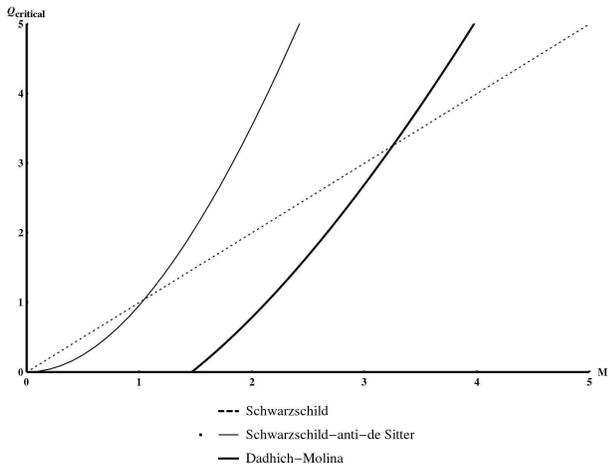}
\caption{Critical charge-mass dependence for different solutions}
\end{center}
\end{figure}

As one can conclude from Fig.3, the DM charge acts in a manner, different from electric one. This statement supports the assert that $q$ in DM solution is rather gravitational charge describing the contribution of additional dimensions.

Concluding this section, we would like to point out that the MD solution generalizes the DM one. We do not perform the study of the MD solution because its behavior is exactly the same as for DM case. For example, increasing number of additional dimensions only shifts a one-horizon curve to a higher position. The unique parameter of the MD solution $\Theta$ is the Weyl's tensor of additional dimensions as the normalization factor. A smaller $\Theta$ defines lowering the one-horizon curve.

\section{Orbital effects}
\label{Orbits}

Orbital effects play an important role in black hole-like solutions, as they are coupled with an accretion picture that is actually observed \cite{Shakura}. We assume cosmological and astronomical scales and, therefore, can use several approximations. We assume that solution has a mass of stellar range or higher, hence one may work with series expansion of a metric. Thus, it is possible to use the asymptotic series \eqref{1}. We work both with $f_+$ and $f_-$ branches, because asymptotically orbital effects depend upon the $M$ and $q$ only.

Because the MD metric is static and spherical symmetric the standard method presented in the Chandrasekhar book \cite{chandrasekhar1998mathematical} can be applied effectively to study stable orbits. Already, this method has been used in \cite{Alexeev} for examining the Dadhich-Rezania solution \cite{Dadhich:2000am} that has many similar properties with the DM solution. Therefore, not repeating here the derivation of an application of the Chandrasekhar method, we rather compare the results of its application to both the DM solution and the Dadhich-Rezania solution. 

If parameter $q$ takes a positive value, the expression \eqref{1} transforms into the Reissner-Nordstr{\"o}m
form that describes black hole with electric charge, although $q$ is not related with electric charge. If $q$ has a negative value, the expression \eqref{1} transforms into the aforementioned Dadhich-Rezania solution \cite{Dadhich:2000am} that describes a black-hole in the Randall-Sundrum II model and has the form:
\begin{eqnarray}
ds^2 = f_\text{DR} dt^2 - \cfrac{dr^2}{f_\text{DR}} -r^2 d\Omega^2 ~, \\
f_\text{DR}= 1 -\cfrac{2M}{r}+\cfrac{\beta}{r^2} ~,
\end{eqnarray}
where $M$ is solution's mass, $\beta$ describes bulk influence. The Dadhich-Rezania solution and its properties were studied in \cite{Alexeev, Zakharov:2014lqa, Chirenti_12, Pugliese_11}. As was established in \cite{Alexeev}, the critical dependence $\beta$ upon mass exists and has the form $\beta \sim M^2$. The Dadhich-Rezania solution turns to the Schwarzschild one if $\beta$ is sufficiently smaller than squared solution mass. It was shown that the critical dependence of $\beta$ upon mass has the form:
\begin{equation}\label{beta_constr}
\beta \leq 9/8 M^2 .
\end{equation}
If $\beta$ is bigger than $9/8 M^2$, test particles orbits change \cite{ Pugliese_11} as the quasinormal modes for the scattering \cite{Chirenti_12} and the shadow size of the black hole \cite{Zakharov:2014lqa}.

It is possible to adopt the limitation \eqref{beta_constr} to the DM solution and to obtain the same one for $q$, which would define an area where orbital effects scattering effects and shadow size is different from GR case:
\begin{equation}\label{q_limitation}
\abs{q} \leq \cfrac{9}{8\sqrt{6}} M^2.
\end{equation}
Engaging the recent data on M87 galaxy \cite{Doeleman_12}, we see that there is no evidence in Dadhich-Rezania model manifestation, so the real difference between the DM solution and the Schwarchild one is negligible small for existing accuracy level.
It is possible to demonstrate that the DM solution fits \eqref{q_limitation}. For $f_+$ branch the parameter $q$ can not take values bigger than $-\cfrac18 (6 \abs{M})^{4/3}$. Hence, for large masses there is an area of small $q$ values. The $q$ existence influences the mentioned effects. Further, for big masses charge values are also limited from above. Asymptotically this limitation acts like $q< M^{4/3}$. So, any allowed DM black hole with positive charge has the properties similar to ones in GR. Nevertheless, restriction \eqref{q_limitation} allows to exclude the area with large negative values of $q$.

\section{Thermodynamics and Evaporation}
\label{Thermodynamics}

Hawking radiation is the other important effect that provides an instrument for model testing at cosmological scales. We use both the Hawking method \cite{Hawking:1974sw} and the Shankaranarayanan-Padmanabhan-Srinivasan (SPS) one \cite{Shankaranarayanan:2000qv, Srinivasan:1998ty}. The SPS approach operates with semiclassical wave-functions and describes black hole evaporation in terms of quantum tunneling. The method gives the DM temperature in the following form:
\begin{equation}\label{T_BH}
T=\cfrac{1}{ 4\pi \operatorname{res} \cfrac{1}{f_-}}~,
\end{equation}
where $\operatorname{res}$ is the residue of $f_-$ function on the horizon. The Hawking method gives expression for temperature in terms of surface gravity:
\begin{equation}
T=\cfrac{k}{2\pi},
\end{equation}
where $k$ is surface gravity. It is defined as a coefficient in $k^\mu \nabla_\mu k^\nu =k k^\nu$, where $k^\mu$ is normalized Killing vector that is normal to horizon, $\nabla_\mu$ is a covariant derivative constructed from the dynamical metric.

We performed numerical calculations for $q=0$ and both for the Hawking and the SPS method. Both strategies lead to similar results and provide temperature-mass dependence shown on the Fig. 4.

\begin{center}
\begin{figure}[h]
\includegraphics[width=80mm]{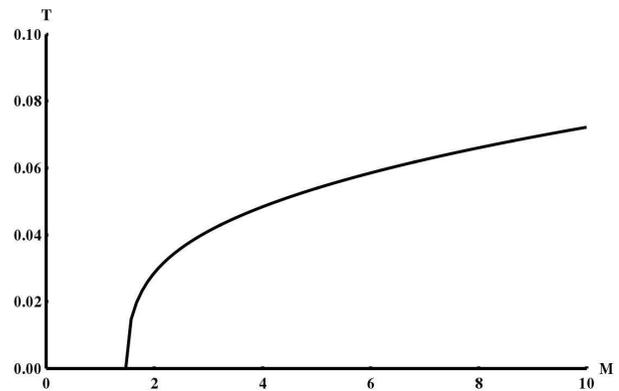}
\caption{Temprature-mass dependence for Dadhich-Molina black hole}
\end{figure}
\end{center}

An interesting particularity of the discussed model is that the temperature grows with $M$ parameter increasing. Similar behavior takes place in Schwarzschild-AdS-like solution \cite{Crisostomo:2000bb}, but for sufficiently big masses only. If the mass is small, the temperature of the Schwarzschild-AdS-like black hole decreases with increasing mass. Note that there is the temperature area that is unreachable for black holes. However, the DM solution temperature only increases with increasing mass, the temperature can achieve arbitrary values.

Numerical calculations show that the decreasing of $q$ corresponds to the temperature growth. In GR black holes with electrical charge have the same temperature-charge dependence. Strictly speaking, $q$ appears to change during the evaporation because properties of outgoing thermal radiation are defined by horizon geometry only. In contrast, it is crucially  to point out that $q$ must not change during evaporation, because it describes the influence of additional dimensions, and their  properties are independent of the 4D black hole local topology.

Within the discussed assumption we estimate that the lifetime of a black hole is infinite. This effect appears because of the temperature decreasing with the DM mass decreasing, which means that the evaporation rate becomes smaller during the evaporation. This conclusion is the opposite to the usual Hawking evaporation model. In addition, the temperature also depends upon $q$ and evaporation rate of black holes with different $q$ values strongly differs. For instance, all black holes on the single horizon curve \eqref{one_horison_curve} do not evaporate.

In this context, the question about the final state of a black hole evaporation arises. One can see that within assumption on $q$ behavior during evaporation of DM black holes with $q>-1$ takes a configuration from one horizon curve \eqref{one_horison_curve} as final state. All these configurations have non zero masses, not depending upon the initial mass of a black hole. In order to obtain the final value of mass for the evaporating DM black hole one should solve Eqs. \eqref{one_horison_curve} with respect to $M$, where $q$ is the initial DM black hole charge. Moreover, for $q>-1$ any final state is a black hole one. For DM black holes with initial $q$ smaller than $-1$ the final states of evaporation are black holes with zero $M$ parameter and non zero $q$ value.

One could conclude that such a situation leads to a contradiction, as in final state the DM black hole has zero mass. This is not true. Calculations of mass from section III are performed with respect to infinite distant observer. In reality one observes a black hole from huge, but finite distance. Hence the expression for mass, measuring by the Earth observer, contains the terms depending on $q$ and $1/L$, where $L$ is distance to a black hole. Therefore the final state of the DM black holes with $q<-1$ has non zero mass. Within the assumption of $q$ conservation during the evaporation one comes to a conclusion that such a state belongs to the DM black hole one. In order to obtain the precise value one should accurately calculate the evaporation rate, taking into account backreaction. We do not do this here because the endpoint masses of the evaporating black holes with $q<-1$ are too small to be found in observations.

Summarizing the results of this section, we conclude that the evaporation strongly depends upon the initial charge $q$. Until one estimates the charge of the DM black hole, it is impossible to find the black holes endpoint mass after the evaporation. Similar, one has no right to make any conclusion about the evaporation rate of the DM black hole until $q$ parameter is unknown.

\section{Discussion and Conclusions}\label{DiscussionConclusions}

In this work we studied Maeda-Dadhich solution obtained in the framework of $N>4$ Einstein-Gauss-Bonnet gravity. We proved the solution stability under linear perturbations, therefore, Maeda-Dadhich metrics is valid for the description of real astrophysical black holes. We investigated the existence of the solution in positive mass region that was not mentioned in the original papers \cite{MaedaDadhich1, Maeda:2006hj, MaedaDadhich3, Molina:2008kh}. We found additional limitations on solution parameters (Section V) that resulted in valuable consequences. The positive branch is valid only for a naked singularity description, while the negative one could describe a black hole with one or two horizons. We established that the black hole temperature is not governed by Hawking evaporation law and this fact causes additional limitations on the Maeda-Dadhich black hole. Unfortunately, the accuracy of existing data on compact objects accretion cannot allow to distinguish thermodynamically stable Maeda-Dadhich black holes from usual Schwarzschild ones. Our result agrees with the estimations obtained earlier \cite{Alexeev, AlexeyevTuryshev}. Moreover, according to the recent results of \cite{Zakharov:2014lqa}, the study of accretion disc shadows and neutral particles scattering cannot help in this regard. Therefore, the Maeda-Dadhich solution can describe astrophysical black holes, but current astronomical data does not allow to distinguish them from Schwarzschild ones from General Relativity.

Concluding, we make two remarks. First, gravitational lensing are considered for testing the presence of both black holes and naked singularities in universe, see \cite{Virbhadra:2002, Virbhadra:2008} and references there in. We do not check such a  possibility here, but plan this in future. Second, it is not secret that EGB gravity is criticized from different sides. So, in \cite{Maldacena+:2014} it is shown that the EGB gravity violates causality on general grounds with the Lagrangian (2.1). The authors \cite{Maldacena+:2014} assert that the only way to avoid this problem is by adding an infinite tower of massive higher-spin particles with a delicate tuning in their couplings. What we describe here is also no the study of the EGB gravity in the general form of the Lagrangian (2.1), but we study it at the Kaluza-Klein split and creating matter by additional dimensions. In fact, we consider ``effective'' GR with the equations \eqref{13}, where, at least, a problem of causality does not appear. Thus, there is no a contradiction with such a criticism.

\section*{Acknowledgments}

This work was partially supported by individual grants from Dmitry Zimin Foundation ``Dynasty'' (S.A. \& B.L.). We are gratefull to Naresh Dadhich for explanation of results in \cite{MaedaDadhich1, Maeda:2006hj, MaedaDadhich3, Molina:2008kh}. We also would like to thank Alexander Zakharov and Alexander Shatskiy for useful discussions. At last, we thank Uliana Voznaya and Deepak Baskaran for the help with improving English.

\appendix

\section{Differential operators for axially symmetric case}\label{sec_DO1}
Here we present the concrete form for differential operators from \eqref{DO_view}.
\begin{widetext}
\begin{eqnarray}
D^{(\nu)}= \cfrac{2\mathfrak{F}'}{r^3}D_{(\theta)} +\cfrac{1}{r^2} \left[ \left\lbrace 4f'(5f-3)+4f r + 3f'r^2 \right\rbrace \cfrac{\pd}{\pd r} + 4f \mathfrak{F} \cfrac{\pd^2}{\pd r^2} \right] ~, \label{DO_axial_begin} \\
D^{(\mu)}=\cfrac{2\mathfrak{F}'}{r^3}D_{(\theta)} -\frac{2}{r^2} \left(\left(r^2-4\right) f''+8 {f'}^2+4 r f'+ f \left(8 f''+2\right)\right) - \\ \nonumber
- \left[ \cfrac{1}{r^2} \left(r^2-4\right) f'+4 f \left(3 f'+r\right) \right] \cfrac{\pd}{\pd r} -\cfrac{4\mathfrak{F}}{r^2 f} \cfrac{\pd^2}{\pd t^2} ~, \\
D^{(\psi)}=\cfrac{2\mathfrak{F}''}{r^2} D_{(\theta)}+\mathfrak{G} \cfrac{\pd}{\pd r} +\cfrac{2 f\mathfrak{F}'}{r} D_{(rt)}  ~,\\
D^{(\lambda)}=-\cfrac{2\mathfrak{F}''}{r^2}\left( \cfrac{\pd}{\pd \theta}- 2 \right)+ \mathfrak{G} \cfrac{\pd}{\pd r} +\cfrac{2 f \mathfrak{F}'}{r} D_{(rt)} ~, \\
D^{(\omega)}=\cfrac{4}{r^2 f} \left[4 \sin^2(\theta ) \left(r f'-1\right)^2-f \left(r^2 {f''}^2+4 {f'}^2+4 r f' \left(f''+\cos (2 \theta )-1\right)+8 \sin^2(\theta )\right)+4 f^2 \sin^2(\theta ) \right] ~, \\
D^{(p)}=\frac{8 (\cos (2 \theta )-3) \left(r f'(r)+f(r)-1\right)^2}{r^4} ~,\\
D^{(q)}=-\cfrac{4}{r^2}\left[ 8 f^2 \sin^2(\theta ) \left(r f'-1\right)+4 f \sin^2(\theta ) \left(r f'-1\right)^2+\left(r f''+2 f'\right)^2+4 f^3 \sin^2(\theta ) \right] ~,\label{DO_axial_end}
\end{eqnarray}
\end{widetext}
where following notations are used
\begin{eqnarray}
D_{(\theta)}= \cfrac{\pd^2}{\pd \theta^2} + \cot\theta \cfrac{\pd}{\pd\theta} ~,\\
D_{(rt)}=\left[\cfrac{\pd^2}{\pd r^2}-\cfrac{1}{ f^2} \cfrac{\pd^2}{\pd t^2} \right] ~,\\
\mathfrak{F}=\cfrac12(4 f+ r^2 -4 ) ~,\\
\mathfrak{G}=\cfrac{2}{r^2} \left[r f' \left(2 f'+r\right)+f \left(4 f'+r \left(2 f''+3\right)\right)\right] ~.
\end{eqnarray}


\begin{thebibliography}{99}

\bibitem{SSM} John H. Schwarz, String Theory and M-Theory. In ``100 Years of Subatomic Physics'', Word Scientific, pp. 519-550 (2013).

\bibitem{EGB-BHs}
S.Alexeyev and M.Ponazanov, Phys.Rev. D {\bf 55}, 2110 (1997);
S. O. Alexeyev, K. A. Rannu, JETP {\bf 114}, 406 (2012)

\bibitem{MaedaDadhich1} H.~Maeda and N.~Dadhich,
Phys.\ Rev. \ D {74}, 021501(R) (2006); hep-th/0605031.

\bibitem{Maeda:2006hj}
  H.~Maeda and N.~Dadhich,
  Phys. Rev. D {\bf 75}, 044007 (2007); hep-th/0611188.

  \bibitem{MaedaDadhich3} N.~Dadhich and H.~Maeda, Int. \ J. \ Mod. \ Phys. \ D
{\bf 17}, 513 (2008); arXiv:0705.2490 [hep-th].

\bibitem{Molina:2008kh}
  A.~Molina and N.~Dadhich,
  Int.\ J.\ Mod.\ Phys.\ D {\bf 18}, 599 (2009);
  arXiv:0804.1194 [gr-qc].

\bibitem{AbbottDeser82}
L.F.~Abbott and S.~Deser,  Nucl. \ Phys. \ B {\bf 195}, 76 (1982).

\bibitem{DT2} S.~Deser and B.~Tekin, Phys. \ Rev. \ D  {\bf 67} 084009 (2003); hep-th/0212292.

\bibitem{DerKatzOgushi} N.~Deruelle, J.~Katz and S.~Ogushi, Class. \ Quantum \ Grav. {\bf 21}, 1971 (2004); gr-qc/0310098.

\bibitem{Petrov2009} A.N.~Petrov, Class. \ Quantum \ Grav. {\bf 26}, 135010 (2009); arXiv:0905.3622 [gr-qc].

\bibitem{Petrov_Lompay_2013_a}	A.N.~Petrov and R.R.~Lompay,
Gen. \ Relat. \ Grav. {|bf 45}, 545 (2013); arXiv:1211.3268 [gr-qc].

\bibitem{PK} A.N.~Petrov and  J.~Katz, Proc. \ R. \ Soc. \ A \ London {\bf 458}, 319 (2002); gr-qc/9911025.

\bibitem{Papapetrou48} A.~Papapetrou, Proc. \ R. \ Irish \ Ac. {\bf 52}, 11 (1948).

\bibitem{Cai+}  R.-G.~Cai, L.-M.~Cao and M.~Ohta,
Phys. \ Rev. \ D {\bf 81} 024018 (2010); arXiv:0911.0245 [hep-th].

\bibitem{chandrasekhar1998mathematical} S.~Chandrasekhar, {\em The Mathematical Theory of Black Holes}, Clarendon Press, 1998.

\bibitem{Dadhich:2000am}
  N.~Dadhich, R.~Maartens, P.~Papadopoulos and V.~Rezania,
  Phys.\ Lett.\ B {\bf 487}, 1 (2000); hep-th/0003061.

\bibitem{Alexeev} S.O.~Alexeev, D.A.~Starodubceva,
  JET {\bf 138}, 652 (2010).

\bibitem{Shakura} Shakura, N.I. and Sunyaev, R.A.,
Astronomy and Astrophysics, {\bf 24}, 337-355, 1973

\bibitem{Zakharov:2014lqa}
  A.~F.~Zakharov,
  Phys.\ Rev.\ D {\bf 90}, no. 6, 062007 (2014)
  [arXiv:1407.7457 [gr-qc]].

\bibitem{Chirenti_12}
C. Chirenti, A. Saa, J. Skakala, Phys. Rev. D {\bf 86}, 124008
(2012).


\bibitem{Pugliese_11}
D. Pugliese, H. Quevedo, R. Ruffini, Phys. Rev. D {\bf 83}, 024021
(2011).

\bibitem{Doeleman_12}
S. S. Doeleman et al., Science {\bf 338}, 355 (2012).

\bibitem{Hawking:1974sw} S.W.~Hawking, 	Commun.\ Math.\ Phys.\ {\bf 43}, 199 (1975)

\bibitem{Srinivasan:1998ty}
  K.~Srinivasan and T.~Padmanabhan,
  Phys.\ Rev.\ D {\bf 60}, 024007 (1999); gr-qc/9812028.

\bibitem{Shankaranarayanan:2000qv} S.~Shankaranarayanan, T.~Padmanabhan and K.~Srinivasan,
  Class.\ Quant.\ Grav.\  {\bf 19}, 2671 (2002); gr-qc/0010042.

\bibitem{Crisostomo:2000bb}
  J.~Crisostomo, R.~Troncoso and J.~Zanelli,
  Phys.\ Rev.\ D {\bf 62}, 084013 (2000); hep-th/0003271.

\bibitem{AlexeyevTuryshev}
Alexeyev S.O., Rannu K.A., Dyadina P.I., Latosh B.N., Turyshev S.G., JETP {\bf 147(5)}, e-Print: arXiv:1501.04217 [gr-qc], accepted for publication (2015)

\bibitem{Virbhadra:2002}
  K.S.~Virbhadra and G.F.R.~Ellis,
  Phys.\ Rev.\ D {\bf 65}, 103004 (2002).

\bibitem{Virbhadra:2008}
  K.S.~Virbhadra and C.R.~Keeton,
  Phys.\ Rev.\ D {\bf 77}, 124014 (2008); arXiv:0710.2333 [gr-qc].

\bibitem{Maldacena+:2014} X.O.~Camanho, J.D.~Edelstein, J.~Maldacena and A.~Zhiboedov, ''Causality Constraints on Corrections to the Graviton Three-Point Coupling'', e-Print: arXiv:1407.5597 [hep-th].

\end{thebibliography}
\end{document}